\documentclass[review]{elsarticle}

\usepackage{lineno,hyperref}
\modulolinenumbers[5]

\journal{Journal of \LaTeX\ Templates}
\usepackage{geometry}
\usepackage{graphicx} 
\usepackage{epstopdf}
\usepackage{indentfirst}
\usepackage{amsmath}
\usepackage{booktabs}
\usepackage{multirow}
\usepackage{amsmath,bm}
\usepackage{multirow}
\usepackage[table,xcdraw]{xcolor}
\begin{document}

\begin{frontmatter}

\title{Perception-oriented Single Image Super-Resolution via Dual Relativistic Average Generative Adversarial Networks}
\author[mymainaddress,mysecondaryaddress]{Yuan Ma}
\author[mymainaddress,mysecondaryaddress]{Kewen Liu}
\author[mythirdaddress]{Hongxia Xiong\corref{mycorrespondingauthor}}
\cortext[mycorrespondingauthor]{Corresponding author}
\ead{xionghongxia@whut.edu.cn}
\author[mymainaddress,mysecondaryaddress]{Panpan Fang}
\author[mymainaddress,mysecondaryaddress]{Xiaojun Li}
\author[mymainaddress,mysecondaryaddress]{Yalei Chen}
\author[myfourtharyaddress]{Chaoyang Liu}
\address[mymainaddress]{School of Information Engineering, Wuhan University of Technology, Wuhan 430070, China}
\address[mysecondaryaddress]{Hubei Key Laboratory of Broadband Wireless Communication and Sensor Networks, Wuhan University of Technology, Wuhan 430070, China}
\address[mythirdaddress]{School of Civil Engineering \& Architecture, Wuhan University of Technology, Wuhan 430070, China}
\address[myfourtharyaddress]{State Key Laboratory of Magnetic Resonance and Atomic Molecular Physics, Wuhan Institute of Physics and Mathematics, Chinese Academy of Sciences, Wuhan 430071, China}

\begin{abstract}
The presence of residual and dense neural networks which greatly promotes the development of image Super-Resolution(SR) have witnessed a lot of impressive results. Depending on our observation, although more layers and connections could always improve performance, the increase of model parameters is not conducive to launch application of SR algorithms. Furthermore, algorithms supervised by $L_1/L_2$ loss can achieve considerable performance on traditional metrics such as PSNR and SSIM, yet resulting in blurry and over-smoothed outputs without sufficient high-frequency details, namely low perceptual index(PI). Regarding the issues, this paper develops a perception-oriented single image SR algorithm via dual relativistic average generative adversarial networks. In the generator part, a novel residual channel attention block is proposed to recalibrate significance of specific channels, further increasing feature expression capabilities. Parameters of convolutional layers within each block are shared to expand receptive fields while maintain the amount of tunable parameters unchanged. The feature maps are subsampled using sub-pixel convolution to obtain reconstructed high-resolution images. The discriminator part consists of two relativistic average discriminators that work in pixel domain and feature domain, respectively, fully exploiting the prior that half of data in a mini-batch are fake. Different weighted combinations of perceptual loss and adversarial loss are utilized to supervise the generator to equilibrate perceptual quality and objective results. Experimental results and ablation studies show that our proposed algorithm can rival state-of-the-art SR algorithms, both perceptually(PI-minimization) and objectively(PSNR-maximization) with fewer parameters.
\end{abstract}
\end{frontmatter}


\section{Introduction}
\setlength{\parindent}{2em}
\indent With the rapid development of artificial intelligence, image Super-Resolution(SR) technology has been widely used in the fields such as smart cities and medical imaging, and has become a research hotspot in computer vision and image processing\cite{park2003super}\cite{protter2008generalizing}. Single image SR refers reconstructing corresponding high-resolution(HR) image according to its low-resolution(LR) counterpart\cite{dong2015image}.

\indent According to different principles, image SR algorithms can be divided into interpolation-based, reconstruction-based and learning-based algorithms. In this paper, we mainly focus on SR algorithms based on convolutional neural networks(CNN) or generative adversarial networks(GAN)\cite{dong2015image}\cite{ledig2017photo}.

\indent Dong et al. first propose the pioneer CNN-based algorithm SRCNN\cite{dong2015image}, which can be divided into three stages, namely feature extraction, feature nonlinear mapping, and upsampling reconstruction to actualize end-to-end learning. CNN-based algorithms no longer explicitly learn external dictionaries, but implicitly learn kernel parameters of middle layers, which have better expression abilities than former algorithms.

\indent To alleviate the problem that deep neural networks are vulnerable to suffer from gradient vanishing and network degradation problems, He and Huang propose ResNet(Residual neural network) and DenseNet(Densely connected neural network), respectively\cite{he2016deep}\cite{huang2017densely}. Ledig et al. propose SRResNet and SRGAN based on ResNet and original GAN. SRResNet introduces abundant global and local skip connections, so that the majority of low-frequency contents can be directly transmitted to the very end of the network through skip connections\cite{ledig2017photo}. Tong et al. propose SRDenseNet based on DenseNet, with which bring the advantages of enhancing feature propagation\cite{tong2017image}. Based on the observation that stacking more layers or adding more connections could always improve performance, the increase of model parameters is not conducive to apply SR algorithms to industry, leaving them staying in academia\cite{zhang2018residual}\cite{zhang2018image}.

\indent Attention mechanism refers to neural network focusing on certain channels or regions. According to different interests, it can be divided into channel attention and spatial attention mechanism. Zhang et al. first introduce channel attention mechanism to SR to adaptively rescale channel-wise features and propose very deep residual channel attention networks(RCAN) which consists of several residual groups with long and short skip connections\cite{zhang2018image}. To ease computational complexity, Muqeet et al. propose hybrid residual attention network(HRAN) with cascading hybrid residual attention blocks that effectively integrate multi-scale feature extraction module and channel attention mechanism and global and short skip connections to ease the flow of information without losing important details\cite{muqeet2019hran}. Experimental results show that attention mechanism is capable of increasing feature expression capabilities without substantial increase of the amount of tunable parameters.

\indent Algorithms supervised by $L_1/L_2$ loss or their variants can achieve considerable performance on traditional metrics such as PSNR and SSIM, yet resulting in blurry and over-smoothed outputs without sufficient high-frequency details, causing visually unpleasing. In order to address the problem, Johnson et al. propose perceptual loss by calculating the Euclidean distance of feature maps extracted by VGG-19 through a specific layer\cite{johnson2016perceptual}\cite{simonyan2014very}. Johnson and Ledig et al. applied perceptual loss to the fields of style transferring and image SR, respectively, and achieved great perceptual quality\cite{ledig2017photo}. Wang et al. explore a variant of perceptual loss, namely MINC loss, calculated by a fine-tuned VGG network for material recognition, which focuses on textures rather than object\cite{wang2018esrgan}. However, the gain of perceptual index brought by MINC loss is marginal.

\indent Recently, GAN has aroused great attention from academic and industrial circles since the ability of generating realistic images far exceeds existing algorithms\cite{goodfellow2014generative}. The application of GAN generally follows the same design pattern, define a generator to non-linearly map data from one domain to another, define a discriminator to evaluate the mapping accuracy. Ledig et al. first propose SRGAN based on ResNet and original GAN\cite{ledig2017photo}. Purohit et al. propose MRDN-GAN to achieve perceptually favorably performance\cite{purohit2018scale}\cite{purohit2019mixed}. Both SRGAN and MRDN-GAN consists of a generator part that reconstruct HR images and a discriminator part that discriminate the source of input images. Reconstructed HR images by GAN-based algorithms can achieve considerable perceptual quality, however, original GAN is notorious for its nonconvergence\cite{arjovsky2017wasserstein}\cite{gulrajani2017improved}\cite{jolicoeur2018relativistic}. To fully exploit the prior that half of data in a mini-batch are fake, Wang and Vu et al. both introduce relativistic GAN to SR, propose ESRGAN and PESR and won the first and fourth place of region 3 of PIRM-SR challenge, respectively\cite{wang2018esrgan}\cite{vu2018perception}\cite{blau20182018}.

\indent This paper develops a novel perception-oriented single image SR algorithm via dual relativistic average GAN. Our main contributions are summarized as follows.
\begin{enumerate}[1.]
\item
A novel generator for perception-oriented SR(we call it G-POSR) that consists of cascading residual channel attention blocks for recalibrating significance of specific channels is proposed. Within each block, parameters of convolutional layers are shared to expand receptive fields while maintain the amount of tunable parameters unchanged. Experimental results and ablation studies verified the effectiveness and superiority of the proposed G-POSR. 
\item Inspired by \cite{wang2018esrgan}\cite{vu2018perception}\cite{park2018srfeat}, to fully exploit the prior that half of data in a mini-batch are fake, dual novel discriminators based on relativistic average discriminators that work in pixel domain and feature domain, respectively, are proposed, to direct our G-POSR to generate images with sufficient high-frequency details. Different weighted combinations of perceptual loss and adversarial loss are utilized to supervise the generator to equilibrate perceptual quality and objective results.
\item We fully complied with regulations of PIRM-SR challenge, and compared our algorithm and its variants from the perspective of PSNR-maximization, PI-minimization, parameter amount, etc. with top submissions in each region\cite{blau20182018}. Experimental results and ablation studies show that our proposed algorithm can rival state-of-the-art SR algorithms, both perceptually and objectively with fewer parameters.
\end{enumerate}


\section{Main work}
\subsection{Residual Channel Attention Block}
{\centering\includegraphics[height=3.5cm]{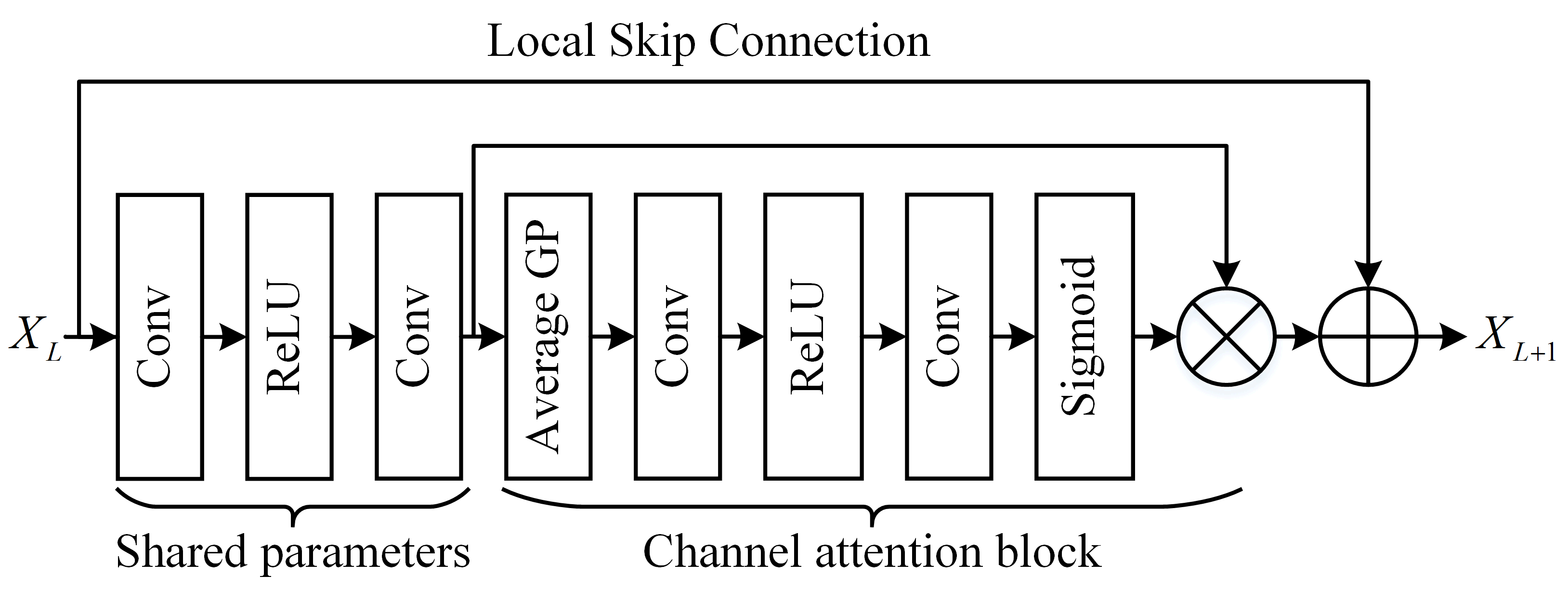}

}
{\centering Fig.1 Proposed residual channel attention block

}

\indent The residual channel attention block is shown in Figure 1 where $ReLU$ and $Sigmoid$ are two different activation functions, symbol $\otimes$ is $Hadamard$ product, symbol $\oplus$ is pixel-wise sum, GP represents global pooling. Each block consists of cascading convolutional layers, activation layer, and channel attention block. Local skip connections are added inside each block. Inspired by \cite{lai2018fast}, we share the parameters of convolutional layers within each block, to expand receptive fields while maintain the amount of tunable parameters unchanged. Specifically, convolutional operations are firstly performed on input feature maps. The size of each kernel is set to $3\times3\times64\times64$, namely 64 kernels whose size are $3\times3$ and the number of channel is 64. After cascading convolutions and activations, the feature maps are fed into channel attention block, and then output to cascading blocks to extract deeper feature representations.\\
\indent In each channel attention block, the dimensions of the input and output feature maps are both $H \times W \times C$. Taking feature maps with dimensions $H\times W \times C$ as input, after average global pooling, two cascading convolutions and activations as Figure 1 shows, corresponding descriptors $\bm{\tau}$ will be obtained.
\begin{equation}
\bm{\tau} = f{\rm{(}}{\bm{W_2}}\delta {\rm{(}}{\bm{W_1}}\bm{x}{\rm{)),}}\tau \in {\bm{R}^{H*W*C}}
\end{equation}
where $\bm{W_1},\bm{W_2}$ represent parameters of the first and second convolutional layer, respectively. The cascading convolutional layers performs channel downscaling and upscaling with ratio 16, after that, corresponding descriptive vector namely descriptors $\bm{\tau}$ for different channels are learned. Smaller descriptive value are adaptively assigned to channels that contain more low-frequency texture contents, this enables network recalibrate significance of specific channels, concentrate more attention on the channels with sufficient high-frequency details. Finally, feature representation through channel attention block can be obtained by multiplying learned descriptor $\bm{\tau}$ and original input. 


\subsection{G-POSR}
{\centering\includegraphics[height=3.5cm]{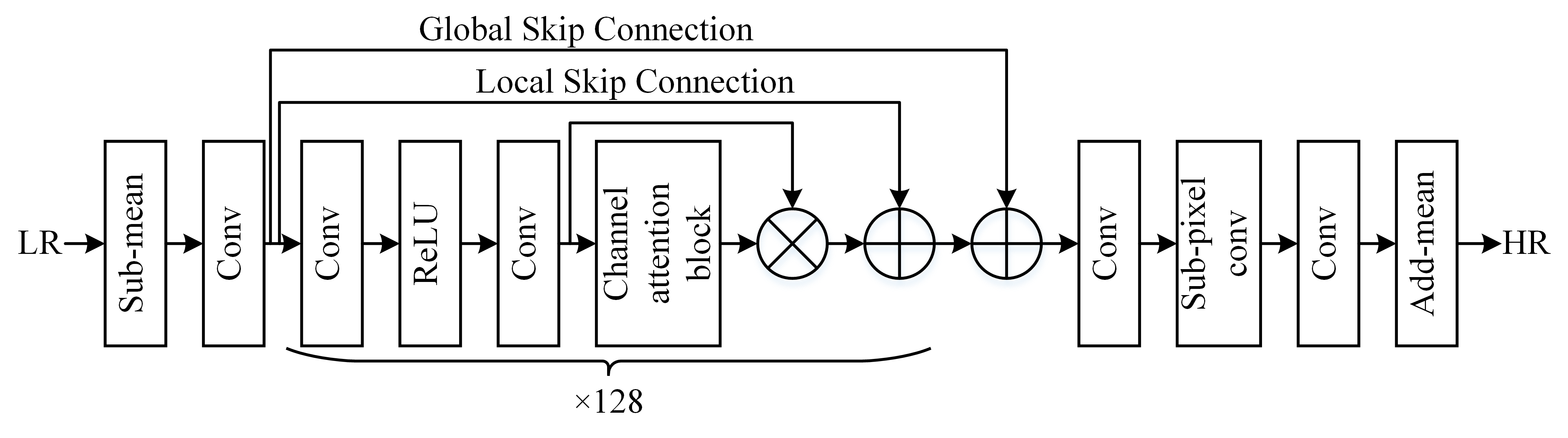}

}
{\centering Fig.2 Network structure of G-POSR

}
\indent The overall network structure of the generator for perception-oriented SR(G-POSR) is shown in Figure 2 which can be divided into three parts, namely feature extraction, nonlinear mapping, and sub-pixel upsampling. The nonlinear mapping part consists of 128 proposed residual channel attention blocks. Local and global skip connections are added within each block and between top and bottom part of the network, greatly alleviating gradient vanishing and network degradation problems. The outputs of nonlinear mapping are then fed into subsequent sub-pixel convolutional layers to obtain final output HR images. Multiple sets of ablation studies according to different block settings are performed, which will be shown in Section 4.


\subsection{Dual Discriminators based on Relativistic Average GAN}
Different from original GAN, to fully exploit the prior that half of data in a mini-batch are fake, relativistic GAN proposed by Alexia introduces the ideology of Turing test and discriminates relativistic authenticity within each mini-batch, that is, determines whether real samples are more real than generated samples. During relativistic GAN's training phase, minimizing generator's loss is equivalent to the minimize $f$-divergence between ${p_{data}}({x_{real}}){p_g}({x_{fake}})$ and ${p_{data}}({x_{fake}}){p_g}({x_{real}})$, where ${p_{data}}$ represents real data distribution, ${p_g}$ represents generated data distribution, instead of minimizing $f$-divergence between ${p_g}$ and ${p_{data}}$ as in original GAN. Relativistic average GAN is a derivative algorithm of relativistic GAN designed to reduce time complexity from $O({N^2})$ to $O(N)$. 

{\centering\includegraphics[height=2.5cm]{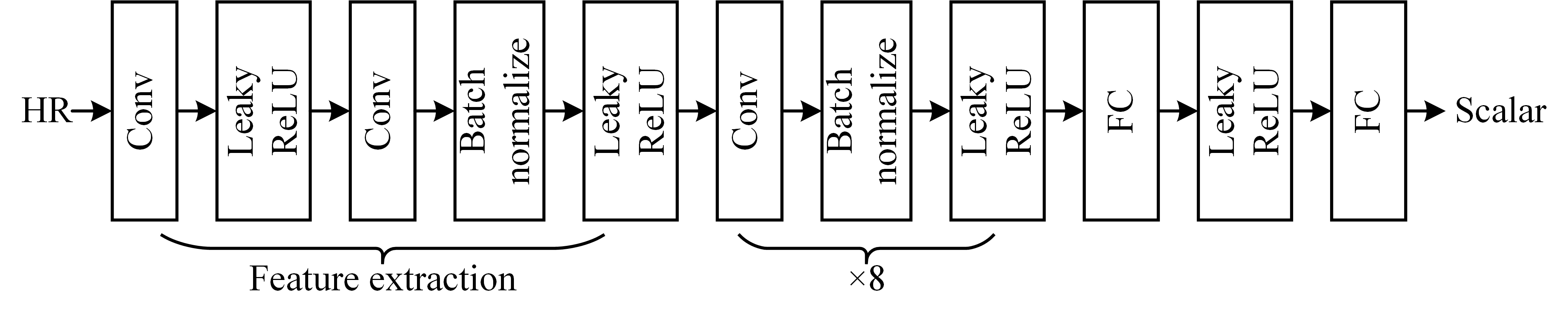}

}
{\centering Fig.3 Network structure of pixel discriminator

}

Dual discriminators work in pixel domain and feature domain respectively. The network structure of the pixel discriminator is shown in Figure 3, where $LeakyReLU$ is a activation function, $negative\_slope$ is set to 0.2, $FC$ are fully connected layers. After feature extraction and 8 cascading nonlinear mapping blocks, images are fed into cascading $FC$ and activation layers to measure whether real HR image ${x_{real}}$ is more realistic than a generated one ${x_{fake}}$.

{\centering\includegraphics[height=2.5cm]{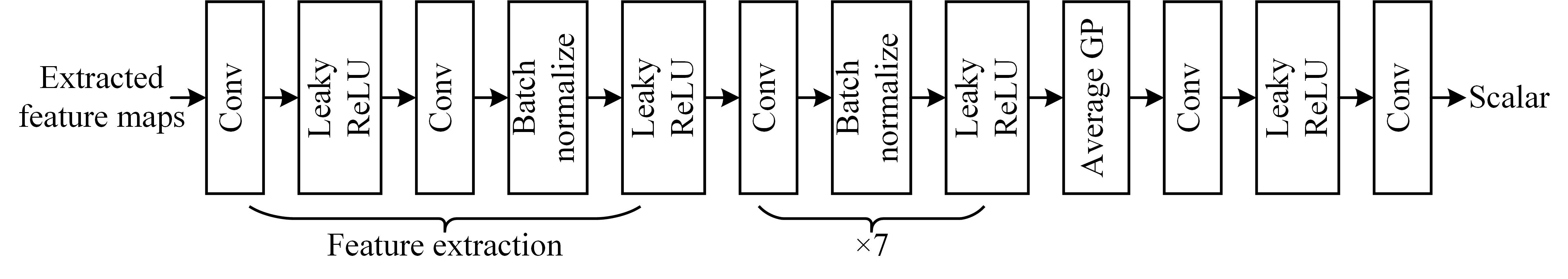}

}
{\centering Fig.4 Network structure of feature discriminator

}

The network structure of the feature discriminator is shown in Figure 4, different from pixel discriminator, we replace FC layers with fully convolutional layers to speed up training and reduce model complexity. After feature extraction and 7 cascading non-linear mapping blocks, feature maps are fed into cascading convolutional and activation layers to measure whether feature maps of real HR image ${\phi}({x_{real})}$ is more realistic than feature maps of a generated one ${\phi}({x_{fake})}$, where ${\phi()}$ represents feature extraction by VGG-19. Inspired by \cite{wang2018esrgan}, we extract feature maps by 4th convolutional layer, before 5th max-pooling layer to avoid over-sparse feature representations.


\subsection{Overall Network Structure}
\indent The overall network structure is shown in Figure 5. The generator can be divided into three parts, namely feature extraction, feature nonlinear mapping, sub-pixel convolution. Mixed generated HR and real HR images are then input to dual discriminators. During training phase, dual discriminators and generator are trained alternately and the loss of the generator and dual discriminators to the generator are fused to direct generator’s training. During testing phase, it's not neccessary to require participation of dual discriminators.

{\centering\includegraphics[height=6cm]{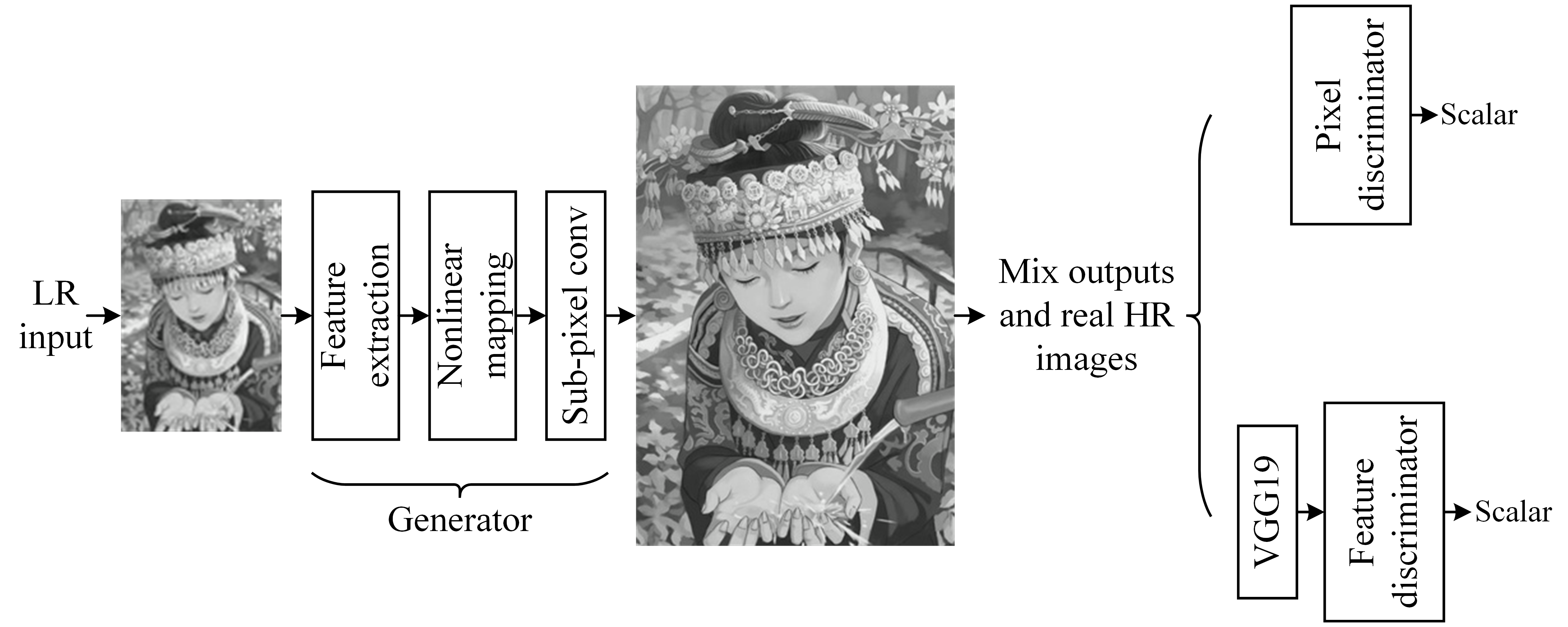}

}
{\centering Fig.5 Overall network structure

}

\subsection{Loss Functions}
\indent Inspired by \cite{wang2018esrgan}, training phase is divided into two stages. In stage one, we use $L_{1-Charbonnier}$ loss proposed in \cite{lai2018fast} to quantify the similarity between generated images and their real HR counterparts. $L_{1-Charbonnier}$ loss function is shown as in equation 2.
\begin{equation}
L_{1-Charbonnier} = \frac{1}{{n*H*W*C}}\sum\limits_{v = 1}^n {\sum\limits_i^W {\sum\limits_j^H {\sum\limits_k^C {\rho (I_{v,i,j,k}^{HR} - I_{v,i,j,k}^{SR})} } } }
\end{equation}
where $I^{SR}$ represents generated images, $I^{HR}$ represents real HR images, $\rho (x) = \sqrt {{x^2} + {\varepsilon ^2}} $, $\varepsilon $ is set to ${10^{ - 3}}$, $H,W,C$ are spatial sizes and channel number of input image, $n$ is number of mini-batch, $I_{v,i,j,k}^{}$ is the pixel value of position $(i,j)$ in ${k_{th}}$ channel of ${v_{th}}$ input image. 

In stage two, we use weighted combinations of perceptual loss, original $L_1$ loss and adversarial loss to supervise training. The adversarial loss of generator of dual discriminators are as follows.
\begin{equation}
L_G^{pixel} = - {E_{{I^{HR}}}}[\log (1 - C_{RaGAN}^{pixel}({I^{HR}},{I^{SR}}))] - {E_{{I^{SR}}}}[\log C_{RaGAN}^{pixel}({I^{SR}},{I^{HR}})]
\end{equation}
\begin{equation}
L_G^{feature} = - {E_{{I^{HR}}}}[\log (1 - C_{RaGAN}^{feature}({I^{HR}},{I^{SR}}))] - {E_{{I^{SR}}}}[\log C_{RaGAN}^{feature}({I^{SR}},{I^{HR}})]
\end{equation}
where
\begin{equation}
C_{RaGAN}^{pixel}({I^{_{HR}}},{I^{_{SR}}}) = \sigma ({D^{pixel}}({I^{_{HR}}}) - {E_{{I^{SR}}}}({D^{pixel}}({I^{SR}})))
\end{equation}
\begin{equation}
C_{RaGAN}^{feature}({I^{_{HR}}},{I^{_{SR}}}) = \sigma ({D^{feature}}({\phi}({I^{_{HR}}})) - {E_{{I^{SR}}}}({D^{feature}}({\phi}({I^{SR}}))))
\end{equation}
where $C$ represents discriminator and $\sigma$ represents $Sigmoid$ function. Thus, the total loss of generator consists three weighted parts as shown in equation 7.
\begin{equation}
{L_G} = {L_{perceptual}} + \lambda {L_1} + {\eta _1}L_G^{pixel} + {\eta _2}L_G^{feature}
\end{equation}
where ${\lambda},{\eta _1},{\eta _2}$ are balance factors to equilibrate perceptual quality and objective results. Experiments of different weighted combinations are performed, which will be shown in Section 4.


\section{Experimental Settings}
\subsection{Training set and test set}
\indent Training images used in this paper are 1-800 images of DIV2K datasets, DIV2K is the designated dataset for NTIRE competition, containing 1000 natural images with a resolution of 2K, of which 1-800 are training sets\cite{Ignatov_2018_ECCV_Workshops}. For objective and perceptual evaluation, algorithms at scaling ratio 4 are tested on public benchmarks Set5, Set14, BSD100, Urban100, Manga109 and PIRM-val. 

\subsection{Parameter Settings and Training Details}
\indent Training HR images are cropped to 2.65 billion sub-images of size $96\times96$ with stide 1 as preprocessing procedure. Corresponding LR images with scaling ratio  4 are obtained by down-sampling using MATLAB bicubic kernel. Data augmentation is performed on training images, which are randomly rotated by ${90^ \circ },{180^ \circ },{270^ \circ }$ and flipped horizontally to obtain more training data. According to GPU memory usage, different mini-batch sizes are set for different scaling ratio. For optimization, proposed algorithms are optimized by ADAM optimizer with ${\beta _1}{\rm{ = }}0.9$, ${\beta _2}{\rm{ = }}0.999$. Total iteration number is set to $2.4 \times {10^7}$. The initial learning rate is set to $5\times{10^{ - 5}}$ and then decreases to half at $[1.44 \times {10^7},4.8 \times {10^6},4.8 \times {10^6}]$ iterations to achieve optimal results. It took nearly two days to train with two GTX 1070Ti.


\section{Experiments and Discussion}
\subsection{Benchmark metrics and details}

For comparisons with state-of-the-art PSNR-oriented algorithms , we test our G-POSR after training phase stage one from three perspectives of PSNR, SSIM and model complexity. For comparisons with state-of-the-art perception-oriented algorithms after training phase stage two, we mainly use perceptual index(PI), one non-reference measurement consisting of Ma’s score and NIQE, first applied in PIRM-SR challenge\cite{ma2017learning}\cite{mittal2012making}. The smaller the PI, the better the perceptual quality. When there is a marginal PI difference (up to 0.01), the algorithm with lower RMSE is better. We fully complied with regulations of PIRM-SR challenge, and compared our algorithm and its variants with top submissions in each region as PIRM-SR divided by RMSE as figure 6 shows\cite{blau20182018}.

{\centering\includegraphics[height=5cm]{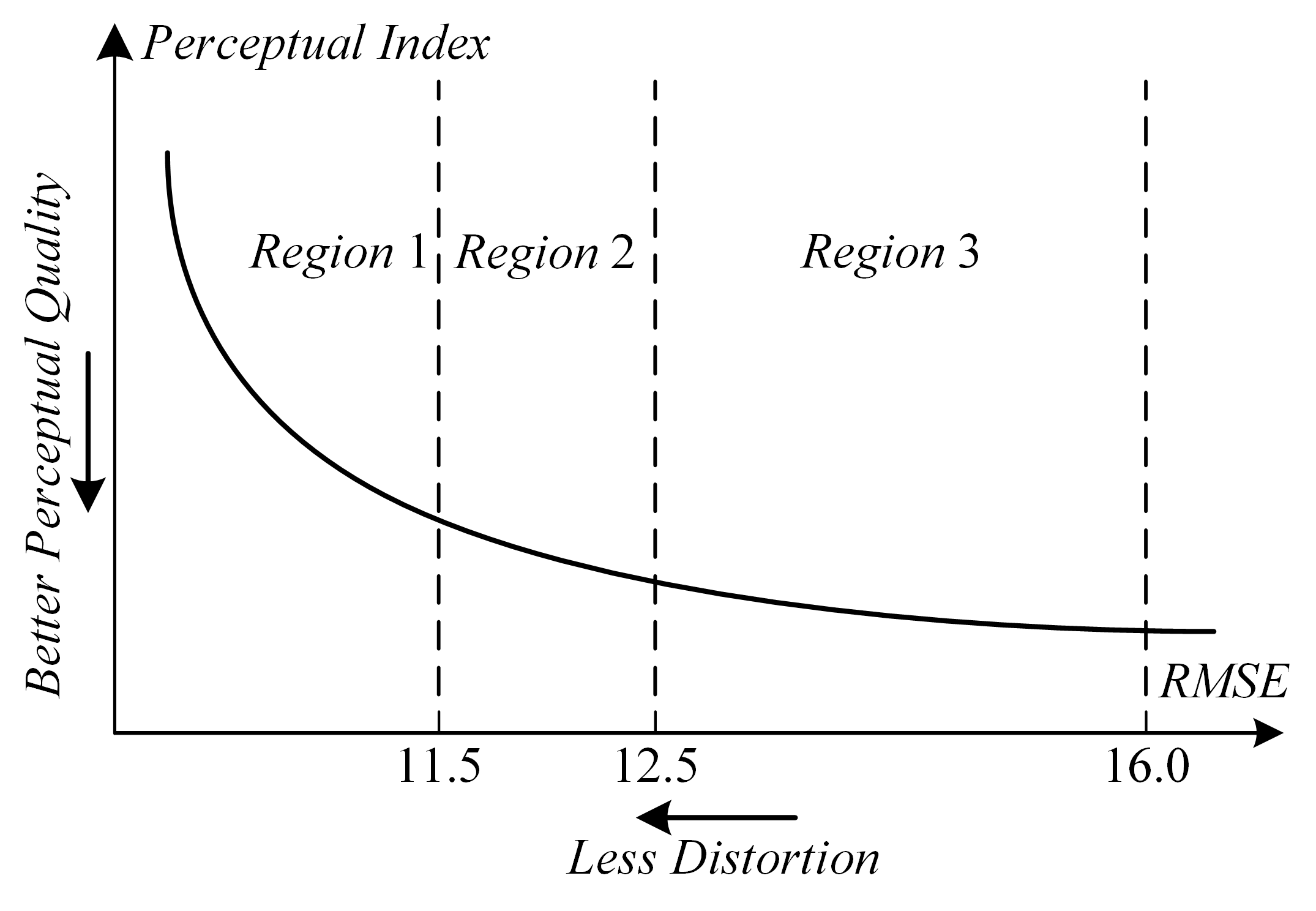}

}
{\centering Fig.6 Perception-distortion plane

}

\subsection{Comparisons with state-of-the-art PSNR-oriented algorithms and discussion}
\indent PSNR and SSIM results of scaling ratio 4 calculated on each testset are averaged and shown in Table 1. The amount of tunable parameters of each algorithm is shown in the rightmost column. When calculating, we first convert the images from RGB to YCbCr color space, remove 4 pixel border and calculate only on Y channel as most algorithms do. Results of comparative algorithms are the best results among their papers and open source project.

{\centering Table 1. Quantitative comparisons with state-of-the-art PSNR-oriented algorithms,\\
maximal values are bold, and second ones are underlined.
\vspace {1em}
\scalebox{0.8}{
\begin{tabular}{c|ccccc|c}
\hline
\multirow{2}{*}{Algorithm} & Set5 & Set14 & BSD100 & Urban100 & Manga109 & \multirow{2}{*}{Parameters} \\
& PSNR/SSIM & PSNR/SSIM & PSNR/SSIM & PSNR/SSIM & PSNR/SSIM & \\ \hline
MSRCAN\cite{cao2019single} & 31.89/0.8907 & 28.33/0.7753 & 27.45/0.7308 & 25.75/0.7733 & 29.71/0.8969 & 4.8M \\
FRSR\cite{soh2019natural} & 32.20/0.8939 & 28.54/0.7808 & 27.60/0.7366 & 26.21/0.7904 & -/- & 4.8M \\
MSRN\cite{li2018multi} & 32.25/0.8958 & 28.63/0.7833 & 27.61/0.7377 & 26.22/0.7905 & 30.57/0.9103 & 6.3M \\
EDSR\cite{lim2017enhanced} & 32.46/0.8968 & 28.80/0.7876 & 27.71/0.7420 & {\underline {26.64/0.8033}} & 31.02/0.9148 & 43M \\
RDN\cite{zhang2018residual} & \textbf{32.47/0.8990} & {\underline {28.81/0.7871}} & \textbf{27.72/0.7419} & 26.61/0.8028 & 31.00/0.9151 & 21.9M \\
D-DBPN\cite{haris2018deep} & {\underline {32.47/0.8980}} & \textbf{28.82/0.7860} & {\underline {27.72/0.7400}} & \textbf{27.08/0.7950} & \textbf{31.50/0.9140} & 10M \\
G-POSR(ours) & 32.44/0.8969 & 28.75/0.7849 & 27.68/0.7389 & 26.44/0.7962 & {\underline {31.09/0.9130}} & 5.1M \\ \hline
\end{tabular}}

}

\indent Clearly from Table 1 that our G-POSR can achieve the best results among the algorithms with parameters fewer than 10M. This demonstrates our algorithm can well equilibrate the amount of parameters and performance. Meanwhile, when comparing with algorithms with a large amount of parameters, such as EDSR, D-DBPN and RDN, there is only a marginal difference, while only needs the 23.7\% and 12\% parameters of RDN and EDSR, respectively. We believe that our algorithm would exceed them if trained with larger patch size or initialized with a smaller initialization than MSRA initialization as in \cite{wang2018esrgan}. Thus, our G-POSR is lightweight and more efficient than comparative state-of-the-art PSNR-oriented algorithms.

\subsection{Comparisons with state-of-the-art perception-oriented algorithms and discussion}
\indent PI results of our perception-oriented SR algorithm with dual discriminators (we call it POSR-GAN) of region 3 of scaling ratio 4 calculated on PIRM-val testset are averaged and shown in Table 2. The amount of tunable parameters of each algorithm is shown in the rightmost column, either. When calculating, we remove 4 pixel border as PIRM-SR challenge request. Results of comparative algorithms are the best results among their papers and open source project. For region 3, due to the RMSE constraints, we empirically set $\lambda$ to 10, set $\eta_1$,$\eta_2$ to 0.125 for training phase stage two after tens of trials.

{\centering Table 2. Quantitative comparisons of region 3 with state-of-the-art perception-oriented algorithms,\\
maximal values are bold, and second ones are underlined.
\vspace {1em}
\scalebox{0.75}{
\begin{tabular}{lllllllll}
\hline
\multicolumn{1}{c}{\multirow{2}{*}{Algorithm}} & \multicolumn{1}{c}{SRGAN\cite{ledig2017photo}} & \multicolumn{1}{c}{CX\cite{mechrez2018learning}} & \multicolumn{1}{c}{EnhanceNet\cite{mechrez2018learning}} & \multicolumn{1}{c}{G-MGBP\cite{mechrez2018learning}} & \multicolumn{1}{c}{PESR\cite{vu2018perception}} & \multicolumn{1}{c}{MRDN-GAN\cite{purohit2018scale}} & \multicolumn{1}{c}{ESRGAN\cite{wang2018esrgan}} & \multicolumn{1}{c}{POSR-GAN(ours)} \\
\multicolumn{1}{c}{} & \multicolumn{1}{c}{PI/RMSE} & \multicolumn{1}{c}{PI/RMSE} & \multicolumn{1}{c}{PI/RMSE} & \multicolumn{1}{c}{PI/RMSE} & \multicolumn{1}{c}{PI/RMSE} & \multicolumn{1}{c}{PI/RMSE} & \multicolumn{1}{c}{PI/RMSE} & \multicolumn{1}{c}{PI/RMSE} \\ \hline
\multicolumn{1}{c}{\textbf{PIRM-val}} & \multicolumn{1}{c}{2.35/-} & \multicolumn{1}{c}{2.133/15.07} & \multicolumn{1}{c}{2.723/15.92} & \multicolumn{1}{c}{2.065/14.32} & \multicolumn{1}{c}{2.013/15.60} & \multicolumn{1}{c}{2.126/14.85} & \multicolumn{1}{c}{\textbf{1.978/15.30}} & \multicolumn{1}{c}{{\underline {2.018/15.02}}} \\ \hline
\multicolumn{1}{c}{Parameters} & \multicolumn{1}{c}{1.5M} & \multicolumn{1}{c}{0.9M} & \multicolumn{1}{c}{0.9M} & \multicolumn{1}{c}{0.3M} & \multicolumn{1}{c}{43.1M} & \multicolumn{1}{c}{5.3M} & \multicolumn{1}{c}{16.7M} & \multicolumn{1}{c}{5.1M}\\ \hline 
\end{tabular}}

}

\indent Clearly from Table 2 that our POSR-GAN can achieve competitive results in contrast to all comparative algorithms. Meanwhile, when comparing with algorithms with a large amount of parameters, such as PESR and ESRGAN, our POSR-GAN outperforms PESR with less distortion and there is only a marginal difference between ESRGAN and our POSR-GAN, while only needs the 11.8\% and 30.5\% parameters of PESR and ESRGAN, respectively. We believe that our algorithm could exceed ESRGAN if post-processed by back projection which can improve PSNR and sometimes lower the PI as in \cite{wang2018esrgan}. Thus, our G-POSR is lightweight and more efficient than comparative state-of-the-art perception-oriented algorithms.

\indent Apart from region3, we also reset $\lambda$,$\eta_1$,$\eta_2$ with different weighted combinations to meet the RMSE constraints of other regions. After tens of trials, for region 1, we empirically set $\lambda$ to 100, set $\eta_1$,$\eta_2$ to 0.005 for training phase stage two. For region 2, we empirically set $\lambda$ to 30, set $\eta_1$,$\eta_2$ to 0.005 for training phase stage two. PI results of our POSR-GAN of region 1 \& 2 of scaling ratio 4 are shown in Table 3 and Table 4. It is worth noting that SRGAN, CX and EnhanceNet are not qualified to region 1 \& 2.

{\centering Table 3. Quantitative comparisons of region 2 with state-of-the-art perception-oriented algorithms,\\
maximal values are bold, and second ones are underlined.
\vspace {1em}
\scalebox{0.75}{
\begin{tabular}{ccccccccc}
\hline
\multirow{2}{*}{Algorithm} & SRGAN\cite{ledig2017photo} & CX\cite{mechrez2018learning} & EnhanceNet\cite{sajjadi2017enhancenet} & G-MGBP\cite{michelini2019multigrid} & PESR\cite{vu2018perception} & MRDN-GAN\cite{purohit2018scale} & ESRGAN\cite{wang2018esrgan} & POSR-GAN(ours) \\
& PI/RMSE & PI/RMSE & PI/RMSE & PI/RMSE & PI/RMSE & PI/RMSE & PI/RMSE & PI/RMSE \\ \hline
\textbf{PIRM-val} & -/- & -/- & -/- & 2.537/12.49 & 2.600/12.42 & 2.760/12.11 & \textbf{2.424/12.50} & {\underline {2.425/12.50}} \\
Parameters & - & - & - & 0.3M & 43.1M & 5.3M & 16.7M & 5.1M \\ \hline
\end{tabular}}

}

{\centering Table 4. Quantitative comparisons of region 1 with state-of-the-art perception-oriented algorithms,\\
maximal values are bold, and second ones are underlined.
\vspace {1em}
\scalebox{0.75}{
\begin{tabular}{lllllllll}
\hline
\multicolumn{1}{c}{\multirow{2}{*}{Algorithm}} & \multicolumn{1}{c}{SRGAN\cite{ledig2017photo}} & \multicolumn{1}{c}{CX\cite{mechrez2018learning}} & \multicolumn{1}{c}{EnhanceNet\cite{mechrez2018learning}} & \multicolumn{1}{c}{G-MGBP\cite{mechrez2018learning}} & \multicolumn{1}{c}{PESR\cite{vu2018perception}} & \multicolumn{1}{c}{MRDN-GAN\cite{purohit2018scale}} & \multicolumn{1}{c}{ESRGAN\cite{wang2018esrgan}} & \multicolumn{1}{c}{POSR-GAN(ours)} \\
\multicolumn{1}{c}{} & \multicolumn{1}{c}{PI/RMSE} & \multicolumn{1}{c}{PI/RMSE} & \multicolumn{1}{c}{PI/RMSE} & \multicolumn{1}{c}{PI/RMSE} & \multicolumn{1}{c}{PI/RMSE} & \multicolumn{1}{c}{PI/RMSE} & \multicolumn{1}{c}{PI/RMSE} & \multicolumn{1}{c}{PI/RMSE} \\ \hline
\multicolumn{1}{c}{\textbf{PIRM-val}} & \multicolumn{1}{c}{-/-} & \multicolumn{1}{c}{-/-} & \multicolumn{1}{c}{-/-} & \multicolumn{1}{c}{3.943/11.49} & \multicolumn{1}{c}{3.321/11.37} & \multicolumn{1}{c}{3.953/11.42} & \multicolumn{1}{c}{\textbf{2.933/11.50}} & \multicolumn{1}{c}{{\underline {3.178/11.48}}} \\
\multicolumn{1}{c}{Parameters} & \multicolumn{1}{c}{-} & \multicolumn{1}{c}{-} & \multicolumn{1}{c}{-} & \multicolumn{1}{c}{0.3M} & \multicolumn{1}{c}{43.1M} & \multicolumn{1}{c}{5.3M} & \multicolumn{1}{c}{16.7M} & \multicolumn{1}{c}{5.1M} \\ \hline 
\end{tabular}}

}

\indent Clearly from Table 3 \& 4, in region 2, our POSR-GAN can achieve competitive results in contrast to all comparative algorithms with lower PI, less distortion, fewer parameters and there is only a marginal difference (0.001 on PI) between ESRGAN and our POSR-GAN. This demonstrates our algorithm has superior capacity of generalization to trade-off between perceptual quality and objective results.

However, in region 1, although our POSR-GAN achieve competitive results in contrast to comparative algorithms except ESRGAN, there is a non-negligible difference (0.245 on PI) between ESRGAN and our POSR-GAN. We believe that our algorithm could perform better after more trials of different weighted combinations. We will show qualitative results in subsection 4.5.

\subsection{Ablation studies and discussion}
\indent In order to verify the effectiveness of each component of our G-POSR , ablation studies are performed to compare their differences. Results are shown in Table 5 and detailed discussions are followed.

{\centering Table 5. Ablation studies results of G-POSR, \\
training details of all variant algorithms are as same as subsection 3.2.
\vspace {1em}
\scalebox{0.8}{
\begin{tabular}{cccc|cc}
\hline
\multicolumn{4}{c|}{Settings} & \multicolumn{2}{c}{Results} \\ \hline
Num of blocks & Num of channels & Channel attention block & Parameter shared & PSNR on Set5 x4 & Parameters \\ \hline
32 & 64 & $\surd$ & $\surd$ & 31.90 & 1.54M \\
64 & 64 & $\surd$ & $\surd$ & 32.13 & 2.74M \\
128 & 16 & $\surd$ & $\surd$ & 30.93 & 0.33M \\
128 & 32 & $\surd$ & $\surd$ & 31.74 & 1.29M \\
128 & 64 & $\surd$ & $\times$ & \underline{32.39} & 9.86M \\
128 & 64 & $\times$ & $\surd$ & 32.28 & 5.06M \\
128 & 64 & $\surd$ & $\surd$ & \textbf{32.44} & 5.14M \\ \hline
\end{tabular}}

}

\begin{enumerate}[1.]
\item
\textbf{Channel attention block removal.}
We first remove all the channel attention blocks in our G-POSR. The decrease of the amount of parameters is merely 0.08M, however, we notice 0.16dB drop of PSNR on Set5 when scaling ratio is 4. This demonstrates the effectiveness of channel attention block in each residual block, achieving non-negligible PSNR promotion with little complexity cost.
\item
\textbf{Parameters not shared.}
As we do not share the parameters of convolutional layers within each block, the amount of our G-POSR increases to 9.86M. However, we notice 0.05dB drop of PSNR on Set5 when scaling ratio is 4. Based on our observation, deeper network with sufficient connections could always improve performance. However, during training phase, more connections (e.g. channel concatenation in last layer of DenseNet) always means more occupied GPU memory and lower GPU utilization, requiring more training time. The increase of training time is also not conducive to launch application of SR algorithms. This's why we abandoned DenseNet as basic network. This verifies the effectiveness of parameter sharing strategy and demonstrates that larger receptive fields are more effective than deeper networks, under same constraints of model complexity. Based on the above assumption, maybe we could try dilated convolutions later rather than simply sharing the parameters.
\item
\textbf{Network depth and width of G-POSR.}
As described in subsection 2.2, multiple sets of experiments according to different block settings are performed. We
gradually modify the number of channels and blocks. Clearly from Table 5, deeper or wider model can further improve performance, as expected. The variant algorithm with 32 blocks and 64 channels achieve competitive results on Set5 with MSRCAN, while only needs 32\% parameters of MSRCAN. This fully proves the superior flexibility and generalization ability of our algorithm. According to needs of different application scenario, derivative algorithms can be selected to satisfy needs of different users or different platforms.
\end{enumerate}

\subsection{Qualitative results and discussion}
In order to evaluate perceptual performance of our G-POSR and POSR-GAN, four sets of test images with rich textures and complex details are selected for comparison, which are $Baby$ from Set5, $Lenna$ from Set14, \textit{43067} from B100 and \textit{image001} from Urban100, as shown in figure 7. The suffix '-R3' represents current image belongs to region 3, and so forth. In general, PSNR-oriented algorithms yield over-smoothed results comparing with perception-oriented algorithms. For results of $Baby$ from Set5, MSRCAN, MSRN and our G-POSR fail to recover texture details in eyelashes and eyeballs. In general, perception-oriented algorithms can produce more convincing results. G-MGBP yield promising results, but there are some unpleasant green dots in it (please zoom in to see the details). PESR and MRDN-GAN produce a clear but smooth image which is very close to the HR, losing high frequency details in the eyeball. Our proposed POSR-GAN-R3 produce faithful images with sharper edges and more detailed textures, and its variants can well equilibrate perceptual quality and objective results, with fewer parameters. For results of $43074$ from B100, PSNR-oriented algorithms yield over-smoothed results that one cannot distinguish what the image contains. Perception-oriented algorithms generate sharper edges and more detailed textures but fail to show the sense of depth of the image. Our proposed POSR-GAN-R3 produce faithful results relatively that one can at least distinguish edges and directions of feathers. 

Qualitative results verified the effectiveness and superiority of our proposed POSR-GAN. All test results will be released at
https://github.com/ascetic-yuanma/POSR-GAN.

{\centering\includegraphics[height=18.2cm]{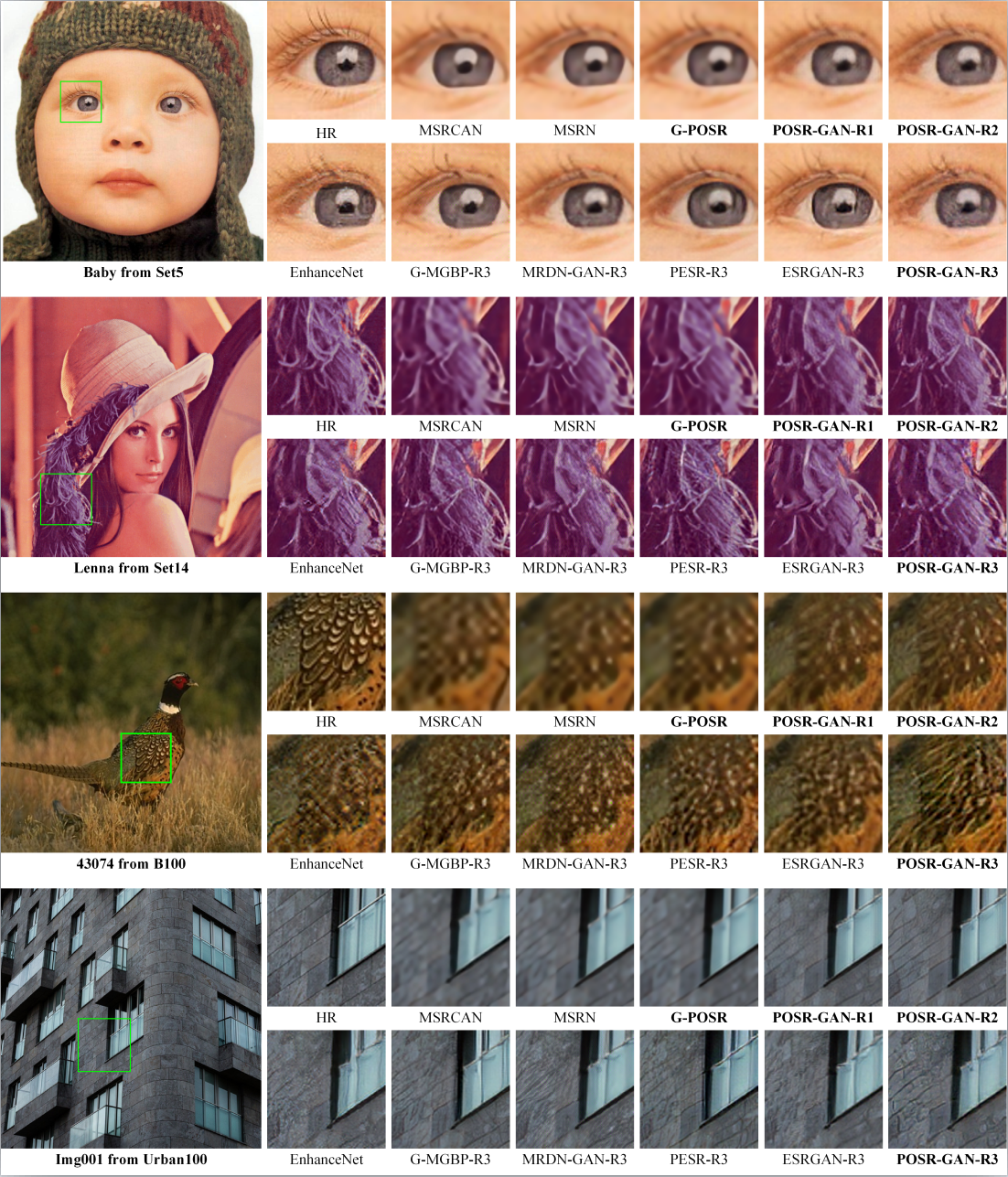}

}
{\centering Fig.7 Comparison of qualitative results of each algorithm when scaling ratio is 4, zoom in to see the details.

}

\section{Conclusion}
In this paper, we have proposed a novel perception-oriented single image SR algorithm via dual relativistic average GAN. We have formulated a novel residual channel attention block, further increasing feature expression capabilities. In addition, parameters sharing are used to expand receptive fields while maintain the amount of tunable parameters unchanged. We have also added two relativistic average discriminators, fully exploiting the prior that half of data in a mini-batch are fake. Moreover, different weighted combinations of perceptual loss and adversarial loss are utilized to equilibrate perceptual quality and objective results. Experimental results and ablation studies verified the effectiveness and superiority of the proposed algorithm, and show that our proposed algorithm can rival state-of-the-art SR algorithms, both perceptually(PI-minimization) and objectively(PSNR-maximization) with fewer parameters.


\section*{Acknowledgments}
This work was supported by the National Key R\&D Program of China (2018YFC0115000). 

{
\bibliography{elsarticle-template}

\begin{thebibliography}{10}
\expandafter\ifx\csname url\endcsname\relax
  \def\url#1{\texttt{#1}}\fi
\expandafter\ifx\csname urlprefix\endcsname\relax\def\urlprefix{URL }\fi
\expandafter\ifx\csname href\endcsname\relax
  \def\href#1#2{#2} \def\path#1{#1}\fi

\bibitem{park2003super}
S.~C. Park, M.~K. Park, M.~G. Kang, Super-resolution image reconstruction: a
  technical overview, IEEE signal processing magazine 20~(3) (2003) 21--36.

\bibitem{protter2008generalizing}
M.~Protter, M.~Elad, H.~Takeda, P.~Milanfar, Generalizing the nonlocal-means to
  super-resolution reconstruction, IEEE Transactions on image processing 18~(1)
  (2008) 36--51.

\bibitem{dong2015image}
C.~Dong, C.~C. Loy, K.~He, X.~Tang, Image super-resolution using deep
  convolutional networks, IEEE transactions on pattern analysis and machine
  intelligence 38~(2) (2015) 295--307.

\bibitem{ledig2017photo}
C.~Ledig, L.~Theis, F.~Husz{\'a}r, J.~Caballero, A.~Cunningham, A.~Acosta,
  A.~Aitken, A.~Tejani, J.~Totz, Z.~Wang, et~al., Photo-realistic single image
  super-resolution using a generative adversarial network, in: Proceedings of
  the IEEE conference on computer vision and pattern recognition, 2017, pp.
  4681--4690.

\bibitem{he2016deep}
K.~He, X.~Zhang, S.~Ren, J.~Sun, Deep residual learning for image recognition,
  in: Proceedings of the IEEE conference on computer vision and pattern
  recognition, 2016, pp. 770--778.

\bibitem{huang2017densely}
G.~Huang, Z.~Liu, L.~Van Der~Maaten, K.~Q. Weinberger, Densely connected
  convolutional networks, in: Proceedings of the IEEE conference on computer
  vision and pattern recognition, 2017, pp. 4700--4708.

\bibitem{tong2017image}
T.~Tong, G.~Li, X.~Liu, Q.~Gao, Image super-resolution using dense skip
  connections, in: Proceedings of the IEEE International Conference on Computer
  Vision, 2017, pp. 4799--4807.

\bibitem{zhang2018residual}
Y.~Zhang, Y.~Tian, Y.~Kong, B.~Zhong, Y.~Fu, Residual dense network for image
  super-resolution, in: Proceedings of the IEEE conference on computer vision
  and pattern recognition, 2018, pp. 2472--2481.

\bibitem{zhang2018image}
Y.~Zhang, K.~Li, K.~Li, L.~Wang, B.~Zhong, Y.~Fu, Image super-resolution using
  very deep residual channel attention networks, in: Proceedings of the
  European Conference on Computer Vision (ECCV), 2018, pp. 286--301.

\bibitem{muqeet2019hran}
A.~Muqeet, M.~T.~B. Iqbal, S.-H. Bae, Hran: Hybrid residual attention network
  for single image super-resolution, IEEE Access 7 (2019) 137020--137029.

\bibitem{johnson2016perceptual}
J.~Johnson, A.~Alahi, L.~Fei-Fei, Perceptual losses for real-time style
  transfer and super-resolution, in: European conference on computer vision,
  Springer, 2016, pp. 694--711.

\bibitem{simonyan2014very}
K.~Simonyan, A.~Zisserman, Very deep convolutional networks for large-scale
  image recognition, arXiv preprint arXiv:1409.1556 (2014).

\bibitem{wang2018esrgan}
X.~Wang, K.~Yu, S.~Wu, J.~Gu, Y.~Liu, C.~Dong, Y.~Qiao, C.~Change~Loy, Esrgan:
  Enhanced super-resolution generative adversarial networks, in: Proceedings of
  the European Conference on Computer Vision (ECCV), 2018, pp. 0--0.

\bibitem{goodfellow2014generative}
I.~Goodfellow, J.~Pouget-Abadie, M.~Mirza, B.~Xu, D.~Warde-Farley, S.~Ozair,
  A.~Courville, Y.~Bengio, Generative adversarial nets, in: Advances in neural
  information processing systems, 2014, pp. 2672--2680.

\bibitem{purohit2018scale}
K.~Purohit, S.~Mandal, A.~Rajagopalan, Scale-recurrent multi-residual dense
  network for image super-resolution, in: Proceedings of the European
  Conference on Computer Vision (ECCV), 2018, pp. 0--0.

\bibitem{purohit2019mixed}
K.~Purohit, S.~Mandal, A.~Rajagopalan, Mixed-dense connection networks for
  image and video super-resolution, Neurocomputing (2019).

\bibitem{arjovsky2017wasserstein}
M.~Arjovsky, S.~Chintala, L.~Bottou, Wasserstein gan, arXiv preprint
  arXiv:1701.07875 (2017).

\bibitem{gulrajani2017improved}
I.~Gulrajani, F.~Ahmed, M.~Arjovsky, V.~Dumoulin, A.~C. Courville, Improved
  training of wasserstein gans, in: Advances in neural information processing
  systems, 2017, pp. 5767--5777.

\bibitem{jolicoeur2018relativistic}
A.~Jolicoeur-Martineau, The relativistic discriminator: a key element missing
  from standard gan, arXiv preprint arXiv:1807.00734 (2018).

\bibitem{vu2018perception}
T.~Vu, T.~M. Luu, C.~D. Yoo, Perception-enhanced image super-resolution via
  relativistic generative adversarial networks, in: Proceedings of the European
  Conference on Computer Vision (ECCV), 2018, pp. 0--0.

\bibitem{blau20182018}
Y.~Blau, R.~Mechrez, R.~Timofte, T.~Michaeli, L.~Zelnik-Manor, The 2018 pirm
  challenge on perceptual image super-resolution, in: Proceedings of the
  European Conference on Computer Vision (ECCV), 2018, pp. 0--0.

\bibitem{park2018srfeat}
S.-J. Park, H.~Son, S.~Cho, K.-S. Hong, S.~Lee, Srfeat: Single image
  super-resolution with feature discrimination, in: Proceedings of the European
  Conference on Computer Vision (ECCV), 2018, pp. 439--455.

\bibitem{lai2018fast}
W.-S. Lai, J.-B. Huang, N.~Ahuja, M.-H. Yang, Fast and accurate image
  super-resolution with deep laplacian pyramid networks, IEEE transactions on
  pattern analysis and machine intelligence 41~(11) (2018) 2599--2613.

\bibitem{Ignatov_2018_ECCV_Workshops}
A.~Ignatov, R.~Timofte, et~al., Pirm challenge on perceptual image enhancement
  on smartphones: report, in: European Conference on Computer Vision (ECCV)
  Workshops, 2019.

\bibitem{ma2017learning}
C.~Ma, C.-Y. Yang, X.~Yang, M.-H. Yang, Learning a no-reference quality metric
  for single-image super-resolution, Computer Vision and Image Understanding
  158 (2017) 1--16.

\bibitem{mittal2012making}
A.~Mittal, R.~Soundararajan, A.~C. Bovik, Making a “completely blind” image
  quality analyzer, IEEE Signal Processing Letters 20~(3) (2012) 209--212.

\bibitem{cao2019single}
F.~Cao, H.~Liu, Single image super-resolution via multi-scale residual channel
  attention network, Neurocomputing 358 (2019) 424--436.

\bibitem{soh2019natural}
J.~W. Soh, G.~Y. Park, J.~Jo, N.~I. Cho, Natural and realistic single image
  super-resolution with explicit natural manifold discrimination, in:
  Proceedings of the IEEE Conference on Computer Vision and Pattern
  Recognition, 2019, pp. 8122--8131.

\bibitem{li2018multi}
J.~Li, F.~Fang, K.~Mei, G.~Zhang, Multi-scale residual network for image
  super-resolution, in: Proceedings of the European Conference on Computer
  Vision (ECCV), 2018, pp. 517--532.

\bibitem{lim2017enhanced}
B.~Lim, S.~Son, H.~Kim, S.~Nah, K.~Mu~Lee, Enhanced deep residual networks for
  single image super-resolution, in: Proceedings of the IEEE conference on
  computer vision and pattern recognition workshops, 2017, pp. 136--144.

\bibitem{haris2018deep}
M.~Haris, G.~Shakhnarovich, N.~Ukita, Deep back-projection networks for
  super-resolution, in: Proceedings of the IEEE conference on computer vision
  and pattern recognition, 2018, pp. 1664--1673.

\bibitem{mechrez2018learning}
R.~Mechrez, I.~Talmi, F.~Shama, L.~Zelnik-Manor, Learning to maintain natural
  image statistics, arXiv preprint arXiv:1803.04626 (2018).

\bibitem{sajjadi2017enhancenet}
M.~S. Sajjadi, B.~Scholkopf, M.~Hirsch, Enhancenet: Single image
  super-resolution through automated texture synthesis, in: Proceedings of the
  IEEE International Conference on Computer Vision, 2017, pp. 4491--4500.

\bibitem{michelini2019multigrid}
P.~N. Michelini, H.~Liu, D.~Zhu, Multigrid backprojection super--resolution and
  deep filter visualization, in: Proceedings of the AAAI Conference on
  Artificial Intelligence, Vol.~33, 2019, pp. 4642--4650.

\end{thebibliography}
}

\end{document}